\documentclass[12pt]{article}

\usepackage{graphicx}
\usepackage{amsmath,amsthm,amssymb}

\title{ Neutron-Mirror Neutron conversion in Vacuum, Trap, Material and Neutron Star \thanks{e3 for v5 (arXiv), -- 04-07-22}}
\author{  \textbf{B.O. Kerbikov} \thanks{correspondence: bkerbikov@gmail.com} \bigskip \\
 Lebedev Physical Institute,\\ Moscow 119991, Russia \medskip \\
 Moscow Institute of Physics and Technology, \\ Dolgoprudny 141700, Moscow Region, Russia }

\newcommand{\beq}{\begin{eqnarray}}
 \newcommand{\eeq}{\end{eqnarray}}
\newcommand{\be}{\begin{equation}}
 \newcommand{\ee}{\end{equation}}

\def\ga{\mathrel{\mathpalette\fun >}}
\def\fun#1#2{\lower3.6pt\vbox{\baselineskip0pt\lineskip.9pt
\ialign{$\mathsurround=0pt#1\hfil ##\hfil$\crcr#2\crcr\sim\crcr}}}

\newcommand{{\SD}}{\rm SD}

\newcommand{{\Mc}}{\mathcal{M}}

\newcommand{\vesig}{\mbox{\boldmath${\rm \sigma}$}}

\newcommand{\veR}{\mbox{\boldmath${\rm R}$}}

\newcommand{\veV}{\mbox{\boldmath${\rm V}$}}

\begin{document}
\maketitle
\begin{abstract}
\noindent The possibility of neutron  swapping between the ordinary and  mirror sectors is today a subject of a  great many theoretical and  experimental studies. In  this paper, we  investigate the  neutron-mirror  neutron  transitions in different  environments  from vacuum to neutron star. Our approach is based on the density matrix formalism, Lindblad and Bloch equations and the implication of the seesaw mechanism to the  Hamiltonian diagonalization.

 \bigskip

 \end{abstract}

 \section{Introduction}

The observation of neutron transformation to mirror neutron would be a discovery of fundamental importance. This would demonstrate the existence of a hidden copy of the ordinary particle sector. The idea goes back to Lee and Yang suggestion \cite{1} to compensate the parity violation in weak interactions by the introduction of the right-handed protons. The concept of mirror world as an independent parallel sector took a distinct form in a seminal paper by Kobzarev, Okun and Pomeranchuk \cite{2}. Ordinary and mirror particles can communicate via gravitation  and via oscillations between neutral particles from both sectors [$3-7$] To our knowledge this neutron-mirror neutron $n–n’$ mixing was first considered in \cite{5} where it was also shown that such transitions do not destabilize nuclei. A comprehentive rewiew of the status of the mirror matter concept for 50 years 1956-2006 was given in [$8$]. The review of the early searches for neutron to mirror neutron convertion was presented in \cite{9}. At present mirror matter, and in particular $n–n’$ transformation, is in the focus of intense theoretical and experimental studies. It is not in the scope of the present work to give a review of a great number of relevant publications. An up-to-date situation with regard to performed and planning experiments and a comprehentive list of references may be found in \cite{10}. The most important among the planned experiments is the HIBEAM/NNBAR \cite{10} to be performed at the European Spalation Source \cite{10}. Two significant experiments, namely STEREO \cite{11} and MURMUR \cite{12}, deserve a special mention. These are passing-through-wall experiments with nuclear reactor acting as a source of neutrons/hidden neutrons. This experimental strategy is similar to short-baseline search for sterile neutrinos [$13-17$]. 

The present limit for a free $n–n’$ oscillation time obtained by \cite{18} is $\tau_{n-n'} \geq 448$ s ($90\%$ CL). This result implies the assumption of compensation the Earth’s magnetic field and the absence of the mirror magnetic field. Results with these conditions relaxed may be found in [$10, 19-22$]. Results of STEREO \cite{11} and MURMUR \cite{12} experiments are presented in terms of the swapping probability $p$ (see below) which is correspondingly equal to $p < 3.1 \cdot 10^{-11}$ ($95\%$ CL) \cite{11}, and $p < 4.0 \cdot 10^{-10}$ ($95\%$ CL) \cite{12}.

An important motivation for the current interest to mirror neutron physics is $n–n’$ transition inside the neutron star (NS) and putative  NS-MNS transition, where MNS is a mixed neutron-mirror neutron star [$23-27$]. Possible cosmological and astrophysical manifestations of mirror particles have been discussed since the early days of the mirror matter concept [$28-32$]. Another reason why $n–n’$ transition is now a point of attraction is the possibility that Dark Matter is a mirror twin of the Standard Model \cite{4,29, 33,34}. Finalizing this brief introductory review necessary to mention a conjecture that $n–n’$ transition can be responsible for the neutron lifetime anomaly \cite{21, 27, 35, 36, 37}. The neutron lifetime is measured in two types of experiments, the bottle and the beam once,  see the references above. There is a $4\sigma$ discrepancy between the results of the two methods with the beam experiment giving the value of the lifetime higher than the bottle one. Whether this discrepancy may be attributed to $n–n’$ oscillations is an open question. According to \cite{35, 37} the most probable reason of discrepancy is the drawback of the beam method. In \cite{27} the upper bound for $n–n’$ mixing parameter is deduced from the binary pulsars data with the conclusion that the obtained limit excludes the possibility to explain the neutron lifetime anomaly.


The aim of the  present work is to describe the neutron-mirror neutron conversion $n–n’$ using the density matrix formalism. The point is that when the two-state quantum system is embedded into the environment this approach is the most adequate one.
Interaction with the environment leads to the distruction of the density matrix off-diagonal elements and thus to the lost of the coherence. The present work relies on the Lindblad \cite{38,39} and Bloch \cite{40} equations for the density matrix  and in one case on the seesaw  diagonalization of the Hamiltonian.
In our view the  cornerstone of this approach is a seminal paper by
 G.Feinberg and S.Weinberg \cite{41} on the conversion of Muonium into Antimuonium. A while later this work served as a basis for numerous investigations of the transitions into the mirror world and a related subject of hidden particles production.
In recent years 
the Lindblad equation  became a standard tool to describe the   multi-level system embedded into the environment. Just two of a great many examples is the investigation of the neutrino oscillations in plasma \cite{42} and heavy quark systems evolution in quark-gluon plasma \cite{43}. Therefore some equations presented below will be  either taken for granted, or supplemented with minimal explanations.
Our work has some overlap with \cite{44,45,46,47} as will be indicated below.
 The  detailed numerical calculations will be given elsewhere.

\section{Density matrix formalism. Lindblad equation}

We consider the time evolution of the $n–n’$ system embedded in the environment  with vacuum being a particular case. The density matrix of this system has the form 
\be
\hat\rho=\left(\begin{array}{cc}\rho_{11}&\rho_{12}\\\rho_{21}&\rho_{22}\end{array}\right) \equiv\left(\begin{array}{cc}\rho_{1}&x+iy\\x-iy&\rho_{2}\end{array}\right),\label{1}\ee
where indices 1 and 2  correspond to the neutron and mirror neutron. The second expression for $\hat\rho$ in 
(\ref{1}) is a convenient form used  previously in \cite{45,46}. The relation between the two forms is 

\be \rho_1=\rho_{11},~~ \rho_2=\rho_{22},~~ x=\frac{1}{2} (\rho_{12}+\rho_{21}),~~
y=\frac{i}{2}(\rho_{21}-\rho_{12}).\label{2}\ee
The time evolution of the density matrix is described by the Von--Neumann--Liouville equation 
\be
\frac{d\hat\rho}{dt}=-i [\hat H, \hat \rho], \label{3}\ee where $\hat H$ is the Hamiltonian of the system. But this is not the whole story if the system under consideration interacts with the environment. In this case the  reduced density matrix  evolution is given by  the Lindblad equation. Prior to  the introduction of this  equation a historical remark is in order. To our knowledge, the first work in which such an equation was written down and used for the calculation of the  Muonium to  Antimuonium conversion  is \cite{41} by G.Feinberg and S.Weinberg mentioned in the Introduction. These authors derived the needed equation from the  physical arguments long before it was formulated in the  general form \cite{38,39} on the mathematical grounds and started to be called Lindblad equation. However, in his later work \cite{48} S.Weinberg  used the term  Lindblad  equation,   made reference to \cite{38,39} and did not  mention \cite{41}.

The Lindblad equation   time evolution of the density matrix of an open system has  the following form  \cite{38,39}
\be
\frac{d\hat\rho}{dt}=-i [H \hat \rho]+L\rho L^+ -\frac{1}{2}\{L^+L, \hat\rho\}, \label{4}\ee
where $L$ is the  Lindblad operator which  should  satisfy certain conditions \cite{49} but is not  known apriori, and $\{...\}$ is an anticommutator. A pedagogical derivation of the Lindblad  equation may  be found in \cite{49}.

 The Hamiltonian in (\ref{4}) is Hermitean. Dissipation arising from the interaction with the environment is described by the two terms on the  right-hand side of (\ref{4}) containing Lindblad operator $L$. This dissipation is called Lindblad decoherence.  In order to take into account the decay widths of the mass eigenstates (beta decay of $n$ and $n'$) one has to consider the non-Hermitian Hamiltonians. Such generalization of the Lindblad equation has been discussed in literature, see e.g. \cite{50}, and reads
 
 \be
\frac{d\hat\rho}{dt}=-i (H \hat \rho-\hat\rho H^+)+ L\rho L^+ -\frac{1}{2}\{L^+L, \hat\rho\}, \label{4a}\ee

For $n-n'$ conversion $H$ and $L$ in (\ref{4}) are \cite{44,51}
\be
H=\left(\begin{array}{cc} -\frac{2\pi}{k} n v  \operatorname{Re} f (0) + E - i \frac{\gamma}{2}& \varepsilon\\ \varepsilon& E'-i\frac{\gamma'}{2}\end{array}\right),\label{5}\ee
\be L=\sqrt{nv}F, ~~ F=\left(\begin{array}{cc} f(\theta)&0\\0&0\end{array}\right).\label{6}\ee

Here $E,E',\gamma$ and $\gamma'$ are the neutron and mirror neutron energies and widths. The energies and  the widths may differ either due to the broken
mirror symmetry or to the different external conditions  like neutron interaction  with  matter and electromagnetic fields. Magnetic fields, both usual and  mirror \cite{19,20}, will not be explicitly introduced into the  equations.  Magnetic field acting in our world is implicitly included into the energy difference $d=E-E'$, while hypothetical  mirror magnetic field  which   may  lead to  intriguing effects \cite{19,52,53}  is beyond the scope of this  work. The amplitude $f(\theta)$ is the  neutron elastic scattering  one, $n$ stands for the number  density of the surrounding matter, $v$ is the mean relative  velocity  between  the neutron and  the matter particles. The term $ \frac{2\pi}{k} n v \operatorname{Re} f(0)$ corresponds  to the energy  shift related to  the forward scattering. The mirror neutron  is considered as a sterile  particle subjected only to the  decay  with the decay constant  $\gamma' $.  With Eqs. (\ref{5}) and (\ref{6}) Eq. (\ref{4a}) yields

\be
 \dot{\rho}_1=- 2\varepsilon y  -(nv\sigma_r+\gamma) \rho_1,\label{7a}\ee

\be
 \dot{\rho}_2=+ 2\varepsilon y  -\gamma' \rho_2,\label{7b}\ee
\be 
\dot{x} =- Mx + (d+K) y - \frac12 (\gamma+\gamma')x,\label{7c}\ee

\be\dot{y} =- My -(d+K) x + \varepsilon (\rho_1-\rho_2 )- \frac12 (\gamma+\gamma') y, \label{7d}\ee
Here $\sigma_r=\sigma_t - \sigma_e$ is  the neutron  reaction cross section.
It will be assumed   that the neutron  cross section is saturated by $s$-wave. The quantities $K$ and $M$ in (\ref{7c},\ref{7d}) stand for 
\be K= -\frac{2\pi}{k} nv \operatorname{Re} f (0),~~M= \frac{2\pi}{k} nv \operatorname{Im} f(0),\label{8}\ee and $d=E-E'$. Finally, we note that  one  can write the same  set of  equations (\ref{7a}-\ref{7d}) taking for granted equation ({23}) of \cite{41}, namely
\be 
\frac{d\hat \rho}{dt} =- i \tilde{H} \hat \rho + i \hat \rho \tilde H^+ + 2 \pi n  v\int d (\cos \theta) \hat F(\theta) \hat \rho \hat F ^*(\theta), \label{9}\ee
with $\hat F(\theta)$ given by (\ref{6}) and where $\tilde H$ differs from (\ref{5}) by omitting the  simbol $\operatorname{Re}$ in the  amplitude  $f(0)$.
 This way  of reasoning has been 
 adopted, e.g., in \cite{45,54}. Still, we wanted to present  the Lindblad  equation in  its original form  since it is  presently  commonly used  in particle physics.
 
 The set of   differential  equations (\ref{7a}-\ref{7d}) do not  allow simple  closed form  solutions \cite{55}. Therefore in the next  sections we shall resort to the approximations  adequate for the given environment.
 
 \section{Conversion in vacuum  and in the trap}

The well known expression  for $n-n'$ oscillations in vacuum can be immediately  received  from the set of equations  (\ref{7a}-\ref{7d}). To this end we use the  truncated  set of equations with $K=M=\gamma=\gamma'=0$. Taking the derivative  of the equation (\ref{7d}) for $\dot{y}$ and making  obvious  substitutions, one  gets
\be 
\ddot{y} + (d^2+4\varepsilon^2) y =0.\label{10}\ee

We consider the transition $n$ to $n'$. Therefore the initial conditions at $t=0$ are $\rho_1 =1, \rho_2=\rho_{12}=\rho_{21} =0.$ Correspondingly according to (\ref{2}) the initial condition for $y$ is $y(0)=0$, and from (\ref{7d}) $ \dot{y}(0)=\varepsilon$. With this initial condition, one obtains

 \be y= \frac{\varepsilon}{{\Omega}}\sin {\Omega} t, ~~\Omega^2=d^2+4\varepsilon^2.\label{11}\ee
 
 Then equation (\ref{7b}) for $\dot \rho_2$ yields the result
\be \rho_2 =\frac{4\varepsilon^2}{\Omega^2}\sin^2\frac{\Omega}{2} t.\label{12}\ee

Next we consider transitions  in a trap. This process for UCN (ultracold neutrons) has been experimentally  studied in a number  of works \cite{56,57}. On the theoretical side  we resort to \cite{44,45,58}. A general remark is in order. Decoherence drastically supresses oscillations if the collision rate  with environment  is much higher than the  oscillation  frequency.  There are two sources of decoherence  in the trap experiments, namely, collisions with the trap walls and  with the  residual  gas  inside the trap. Here we consider only collisions with the walls,  decoherence  due to the presence  of low density gas  inside  the trap was studied in \cite{44}.

Let $\tau_i$  be the time  interval  between $(i-1)$-th and $i$-th collisions  with  the walls. It is  convenient  to introduce a variable $R_z=\rho_1-\rho_2$ which is the $z$-th  component  of the  Bloch 3-vector $\veR$ \cite{40}. We shall return to the discussion of the decoherence due to the collisions with the trap walls within the $R$-matrix formalism at the end of this section. In \cite{45} this variable is called $s$. Following collisions  step by step  one arrives at the  obvious result \cite{44,45}.

\be R_z(\tau_{n^+}) = \prod^n_{k=1}\cos (2\varepsilon \tau_{k^+}).\label{13}\ee

Note that $\varepsilon \tau \ll 1$, where $\tau$ is the average collision time, $\tau =t/n$, $n$ is the  number of collisions  and we assume $n\gg 1$, $ \tau\simeq 0.1 s$, $\varepsilon^{-1}\geq 448 s$ according to \cite{18}.
On account of $2\varepsilon \tau_k \ll 1$ and in the approximation of equal time intervals between the collisions, one can represent (\ref{13}) as
\be R_z \simeq \prod^n_{k=1} \left( 1 - \dfrac{(2\varepsilon \tau)^2}{2} \right) \simeq \left( 1 - \dfrac{2\varepsilon^2 \tau t}{n} \right)^n \simeq  \exp(-2\varepsilon^2 \tau t), \label{14}\ee
which coincides with (32) of \cite{45}. We see that  collisions with the walls  exponentially suppress oscillations.  Transition probability after $n$ collisions is equal to 
\be \rho_2= \dfrac12 (1 - R_z) \simeq\varepsilon^2\tau t, \label{15}\ee
in line with (35) of \cite{45} and with (39) of \cite{58}. It is interesting to note that  one can came to the  same result by   using the evolution equation  for  the Bloch 3-vector $\veR$ defined as 
\be\hat \rho=\frac12 (1+\veR\vesig),\label{19'}\ee
\be \veR =\left(\begin{array}{c}\rho_{12}+\rho_{21}\\-i(\rho_{21}-\rho_{12})\\\rho_1-\rho_2\end{array}\right)=\left(
\begin{array}{c} 2x\\-2y\\\rho_1-\rho_2\end{array}\right).\label{20'}\ee

We shall use the $\veR$-matrix formalism in the form proposed by L.Stodolsky \cite{59}. Direct examination shows that the Lindblad equation in the form (\ref{4a}), or (\ref{7a}-\ref{7d}) is equivalent to
\be
\dot{\veR}=\veV\times \veR -D_T\veR_T-\gamma \veR,\label{21'}\ee
where
\be\veV = \left( \begin{array}{c}2\varepsilon\\O\\d+K\end{array}\right),~~
D_T = \left( \begin{array}{cc}M&O\\O&M\end{array}\right),~~R_T\left( \begin{array}{c}R_x\\R_y\end{array}\right).\label{22'}\ee

The physical meaning of the three terms in (\ref{21'}) is quite different. The contribution $K$ in $V_Z$ corresponds to the energy shift due to the refraction index. Alternatively, it may be considerd as a supplementary ``magnetic field'' along the $Z$-axis \cite{59}. The third term in (\ref{21'}) is rather trivial. It corresponds to the shrinking of the Bloch vector $\veR$ in length. It may be set equal to zero without distortion of the physical picture. The most important is the second term with $D_T$ giving the quantum friction.  It leads to destroying the off-diagonal elements of the density matrix and the lost of coherence. Using the optical theorem the matrix elements $M$ of $D_T$ are related to the total neutron cross section
\be M=\frac{2\pi}{k} nvImf(0) =\frac12 nv\sigma_t.\label{23'}\ee

In order to recover the result (\ref{15})
for the transition in the trap we take the equation (\ref{7d}) for $\dot y$
 in the form 
 \be \dot y =- My  + \varepsilon (\rho_1-\rho_2).\label{16}\ee
 
 Taking the   derivative once more,  one  obtains  the following equation for   $R_z$  
 \be \ddot{R}_z +M\dot{R}_z +4\varepsilon^2 R_z =0.\label{17}\ee
 
 This is  the equation for  oscillator with friction. When  $M\gg 4   \varepsilon$  it leads to the overdamping solution which at ``long'' times $t\gg1/M$ is proportional to:
 
\be R_z\sim \exp \left( -\frac{4\varepsilon^2}{M} t\right),\label{26'}\ee
and we return to (\ref{15}) 
 provided $M=2/\tau$. Therefore the transition probability is 
 \be \rho_2(t) =\frac12[1-\exp(-2\varepsilon^2\tau t)]\simeq \varepsilon^2\tau t\label{27'}\ee in line with \cite{58}. The general problem of the equivalence between the Lindblad equation and the Bloch vector evolution equation is discussed in \cite{60}.

\section{Seesaw mechanism in strong absorption regime} 

Consider now the $n-n'$  conversion in the  media 
 with strong neutron absorption. Within the  density matrix formalism similar problem has been solved in  \cite{44}. Here we turn to  the Hamiltonian  diagonalization method closely following  \cite{61,62,63}. At the end of this section we shall show how the Lindblad equation works in this case. Consider the limiting  case $M\gg{| K|}$ (see (\ref{8})). This means that we neglect the neutron rescattering.  It makes  the use of Lindblad  equation not obligatory. The  quantity $M$ can  be presented as
 \be M= \frac{2\pi}{k} nv \operatorname{Im} f (0) =\frac12 nv\sigma \simeq \Gamma/2, \label{18}\ee
 where $\sigma$ is the total neutron  cross section. The  last  equality in (\ref{18}) may be explained in the following way. The mean free path  of the neutron  is $L=(n\sigma)^{-1}$,  The corresponding propagation time is  $ t=L/v$, so  that $\Gamma \simeq t^{-1} = nv\sigma. $ The Hamiltonian reads
 \be H=\left( \begin{array} {cc} -iM& \varepsilon\\ \varepsilon&\omega\end{array}\right),\label{19} \ee
 where we subtracted the part proportional to the  unity matrix, $\omega =E'-E=-d$. Diagonalization results in the two  eigenvalues
 \be\mu_1 \simeq -iM + \varepsilon^2 \frac{iM}{M^2 +\omega^2}-\varepsilon^2 \frac{\omega}{M^2+\omega^2},\label{20}\ee
 \be\mu_2 \simeq \omega- \varepsilon^2 \frac{iM}{M^2 +\omega^2}
 +\varepsilon^2 \frac{\omega}{M^2+\omega^2}. \label{21}\ee
 For the degenerate $n-n'$ levels $(\omega=0)$ there is  a huge disparity between the  two eigenvalues. This is a typical seesaw  picture \cite{61}. The wave function evolution is described by the  equation \cite{63}
 \be\psi(t) = \left( \frac{H-\mu_2}{\mu_1-\mu_2} e^{-i\mu_1t} +  \frac{H-\mu_1}{\mu_2-\mu_1} e^{-i\mu_2t} \right)\left(\begin{array}{c} a\\b\end{array}\right),\label{22}\ee 
  where a stands for $\psi_n(0)$ and $b$ for $\psi_{n'}(0)$. In the leading order in $\varepsilon$ and assuming $|\varepsilon (\omega+i M)^{-1}|\ll 1$ and $\varepsilon^2(\omega^2+M^2)^
  {-1}ML\ll 1$, one obtains
\be \psi(t)=\left(\begin{array}{c} ae^{-\frac{\Gamma}{2} t}- b\frac{\varepsilon}{\omega + i \frac{\Gamma}{2}}\left(e^{-\frac{\Gamma}{2}t}-e^{-i\omega t}\right)\\ be^{-i\omega t}- a\frac{\varepsilon}{\omega+ i \frac{\Gamma}{2}}\left(e^{-\frac{\Gamma}{2}t}-e^{-i\omega t}\right)\end{array}\right).\label{23}\ee
We are  interested in mirror    neutron production, so that     $a=1, b=0$ and
 \be \psi_{n'}(t)=-\frac{\varepsilon}{\omega+i\frac{\Gamma}{2}}\left(e^{-\frac{\Gamma}{2}t}-e^{-i\omega t}\right).\label{24}\ee

For $|\psi_{n'}(t)|^2 =\rho_2 (t)$ this yields
\be
|\psi_{n'}(t)|^2 =\frac{\varepsilon^2}{\omega^2+\frac{\Gamma^2}{4}}\left(1+e^{-\Gamma t} -2e^{-\frac{\Gamma}{2}t}\cos \omega t\right).\label{25}\ee
Our results (\ref{24})  and (\ref{25}) coincide with those of \cite{62,63,64}. 
The solution (\ref{25}) corresponds to the dominant role of the eigenvalue $\mu_1$ (see (\ref{20})). If we remove assumptions allowing to obtain (\ref{23}) from (\ref{22}), and consider the solution of (\ref{22}) for the times  

\be \frac{1}{M} \ll t \ll \frac{1}{M}\left( \frac{M}{\varepsilon}  \right)^2, \label{29v2}\ee
we arrive to another solution with the dominant role the eigenvalue $\mu_2$ (\ref{21}) provided $\omega = 0$, or very small. Then $\rho_2(t) = |\psi_{n'}(t)|^2$ reads \cite{61,44} 

\be |\psi_{n'}(t)|^2 =\frac{\varepsilon^2}{\omega^2+\frac{\Gamma^2}{4}} e^{-\delta t}, \label{30v2}\ee
where $\delta = \varepsilon^2 \Gamma \left( \omega^2+\frac{\Gamma^2}{4}\right)^{-1}$. This is the limit of ``long'' times or distances close to or exceeding the absorption length $L = vt$, where $v$ is neutron velocity. Formally, the distance (or time) independent factor in (\ref{30v2}) is obtained from (\ref{25}) at $\Gamma t \gg 1$, or $(n \sigma L)\gg1$. The regime (\ref{30v2}) means that at some thickness $L$ of the absorber such that $(n v \sigma)  t\gg 1$ or $(n \sigma L)\gg 1$ the attenuation of the neutron beam due to the $n-n'$ transition becomes almost constant.

One can solve equation (\ref{22}) in the next order in $\varepsilon/M$ and get the solution which incorporates both limiting regimes (\ref{25}) and (\ref{30v2}). It has the following form
\be |\psi_{n'}(t)|^2=\frac{\varepsilon^2}
{\omega^2+\frac{\Gamma^2}{4}}
e^{-\delta t} \left( 1+ e^{-\Gamma''t}- 2e^{-\frac{\Gamma''}{2}t}\cos\omega'' t\right),\label{38}\ee
where
\be \delta = \varepsilon^2 \frac{\Gamma}{\omega^ 2+M^2},~~ \omega_\varepsilon =\varepsilon^2 \frac{\omega}{\omega^2+M^2},\label{39}\ee
\be\Gamma''=\Gamma-2\delta,~~
\omega'' =\omega+2\omega_\varepsilon.\label{40}\ee

To treat the strong absorbtion regime within the Lindblad equation approach one can take equations (\ref{7b}) and (\ref{7d}) in an oversimplified form
\be\dot\rho_2 =2\varepsilon y,\label{43}\ee  
\be \dot y=-My+\varepsilon(\rho_1-\rho_2).\label{44}\ee
Taking the time derivative in (\ref{44}) once more we obtain
\be \frac{d^2y}{dt^2} + M\frac{dy}{dt} + 4\varepsilon^2 y=0.\label{45}\ee

The initial conditions are $y(0)=0, \dot y(0)=\varepsilon$. Solving (\ref{45}) in the same approximation that lead to (\ref{23}), putting the result into (\ref{43}), we obtain
\be 
\rho_2(t) =\frac{\varepsilon^2}{M^2} \left(1-e^{-Mt}\right)^2 =\frac{4\varepsilon^2}{\Gamma^2} \left(1-e^{-\frac{\Gamma}{2}t}\right)^2,\label{46}\ee
which is the same as (\ref{25}) for $\omega=0$.

Striving for an analytical solution in the strong absorption   regime we were forced to make a drastic approximation
$M\gg|K|,$ or $Im ~f(0)\gg |Re~f(0)|.$  Looking into the NIST table of the neutron scattering length \cite{65} one concludes that this is not the most adequate assumption. As noted already the system of Lindblad equations  (\ref{7a}-\ref{7d}) does not allow a transparent solution without approximations. For illustrative purposes we present a solution in a regime ``opposite'' to the previous one, namely with the negligible absorption $|K|\gg M$. Equation (\ref{7b}) and (\ref{7d}) take the form 

\be\dot\rho_2 =2\varepsilon y,\label{46}\ee  

\be \dot y=-Kx+\varepsilon(\rho_1-\rho_2).\label{47}\ee

 Invoking equation (\ref{7c}) for $\dot x$ we arrive at  the following equation for $ \ddot y$

\be \frac{d^2y}{dt^2} +(K^2+4\varepsilon^2) y=0.\label{48}\ee
Solving (\ref{46}-\ref{48}) with the same initial conditions for $y(0)$ and $\dot y(0)$ we obtain

\be 
\rho_2(t) =\frac{4\varepsilon^2}{K^2+4\varepsilon^2} \sin^2 \frac12 \sqrt{K^2+4\varepsilon^2} t.\label{49}\ee

As expected neutron undergoes oscillations with the time-averaging swapping probability \cite{46}
\be 
P=\frac{2\varepsilon^2}{K^2+4\varepsilon^2}.\label{50}\ee

The actual experimental regime most probably corresponds to $|K|\ga M$ \cite{65} so that both these quantities should be kept in (\ref{7a}-\ref{7d}). With a given set of physical parameters a compete solution of  (\ref{7a}-\ref{7d})  is a cumbersome but tractable task.

\section{A toy model of $n-n'$ conversion in neutron stars}

As stated, the system of  equations (\ref{7a}-\ref{7d}) does not allow a simple  analytical solution. There is a physical situation when it is of minor importance  to control the $n-n'$  conversion at every moment. The process might be slow,  long-lasting and the goal is to predict the  final outcome, namely the balance between $n$ and $n'$  components  at asymptotically long time. This is exactly what may  happen in  neutron star which could  gradually  transform into  a mixed star  consisting of  normal neutron  and mirror neutron  components \cite{23,24}. What follows is a preliminary outline of a future work on this problem.

The system (\ref{7a})-(\ref{7d}) cannot be solved in a closed form but can be integrated in time from $0$ to $\infty$ with given initial conditions. This procedure has been previously performed in \cite{41,45}.

We return to (\ref{7a})-(\ref{7d}) and introduce the following notations \be X=\int^\infty_0 dtx,~~ Y=\int^\infty_0dty,~~ P_i=\int^\infty_0 dt\rho_i,~~i=1,2,\label{26}\ee 
\be R=nv\sigma_{r}+\gamma,~~ \Gamma =\frac12(\gamma+\gamma'+2M),~~ \Delta=d+K.\label{27}\ee
This mode of action and similar  notations were first introduced by G.Feinberg and S.Weinberg \cite{41}. The initial conditions read $P_1=1,P_2=X=Y=0$. The set of equations to be  solved are
\be RP_1+2\varepsilon Y =1,\label{28a}\ee 
\be \gamma'P_2 -2\varepsilon Y =0,\label{28b}\ee
\be \Gamma X - \Delta Y =0,\label{28c}\ee
\be \Gamma Y + \Delta X - \varepsilon (P_1-P_2)=0.\label{28d}\ee
The branching ratio  we seek for   is \be Br = \frac{P_2}{P_1+P_2}.\label{29}\ee

$Br$ is easily found from  (\ref{28a}-\ref{28d}) in the approximation $\varepsilon^2\ll \Gamma^2$ with the result
\be Br\simeq \frac{\Gamma}{\gamma'} \frac{2\varepsilon^2}{\Gamma^2 +\Delta ^2}.\label{30}\ee
As might be expected,  we obtained the same result as \cite{41,45}. Minor  difference,  from (\ref{28a}-\ref{28d}) of \cite{41} is because the problem  considered here is  not completely the same  as in  \cite{41}. The next  task would have been  to implement  the neutron star parameters from, e.g.,  \cite{23,24} and to get the  mirror matter  admixture under different conditions. This will be the subject of the  work in preparation.

\section{Conclusions and outlook}

We have considered the neutron transition into the mirror world in  different conditions from vacuum to neutron stars. It is shown that the reduced density matrix  formalism, the Lindblad and Bloch equations are the most efficient tools to solve this problem. The reason is  that the contact  with the  surroundings leads to the distruction of the  density matrix  off-diagonal elements and  consequently the  loss of the  coherence. Important to note that deconherence and the resulting collapse of the  wave function is a phenomenon beyond the standard quantum mechanics based on the Schrodinger equation with optical potential. The correct description was proposed by G.Feinberg and S.Weinberg \cite{41}. Eventually this approach was coined the name Lindblad equation \cite{38,39}. Relation between the optical potential and the Lindblad equation was discussed in \cite{51,67}. A correct way to incorporate the optical potential into the Lindblad equation  presented in \cite{51} is

\be \frac{d\rho}{dt} =-i[H,\rho]-i(W_{opt}\rho-\rho W^*_{opt})+L\rho L^+,\label{41}\ee

\be \text{ where \quad  } W_{opt}=-2\pi\frac{n}{m_*}f(0).\label{42}\ee

\noindent Subject to minor differences, this equation is equivalent to (\ref{7a})-(\ref{7d}) of the present work.

To summarize, we may say that the problem of the $n-n'$ conversion in a trap is completely solved. The transition in the absorbing material has been studied in \cite{11,12,46,66} and the present work. Our approach is close to that used in \cite{12,46}. In some limiting approximations the results basically coincide, like (9) of \cite{46} and (\ref{50}) of the present work. On the other hand, we can not find an immediate correspondence between (24) of \cite{66} and our results. This poses a serious problem to work at in order to provide a clear guidance to the experiment. As for the intriguing problem of neutron-mirror-neutron star transition, it gets a close attention \cite{23,24} but is far from a complete solution. 


\section{Acknowlegments}

This work has been supported by RFBR grant 18-02-40054. The author is thankful to Leo Stodolsky for enlightening materials on the reduced density matrix  formalism and to M. Sarrazin and M. Khlopov for important remarks. I am indebted to M.S. Lukashov, N.P. Igumnova, N.P. Nemtseva  and A.Simovonian for comments at all stages of the work.

  \end{document}